\documentclass[aps, prd, twocolumn, superscriptaddress, nofootinbib, showpacs]{revtex4}
\usepackage{graphicx}
\usepackage{amsmath}
\def\fsl#1{\setbox0=\hbox{$#1$}                 
   \dimen0=\wd0                                 
   \setbox1=\hbox{/} \dimen1=\wd1               
   \ifdim\dimen0>\dimen1                        
      \rlap{\hbox to \dimen0{\hfil/\hfil}}      
      #1                                        
   \else                                        
      \rlap{\hbox to \dimen1{\hfil$#1$\hfil}}   
      /                                         
   \fi}                                         %
\newcommand{\tr}{\mbox{tr}}

\newcommand{\VEV}[1]{\langle #1 \rangle}
\newcommand{\gtrsim}{\mathop{>}\limits_{\displaystyle{\sim}}}
\newcommand{\lessim}{\mathop{<}\limits_{\displaystyle{\sim}}}
\newcommand{\UL}[1]{\underline{#1}}
\makeatletter
\@addtoreset{equation}{section}
\makeatother

\begin{document}
\title{
Does the three site Higgsless model survive
the electroweak precision tests at loop?
}  
\date{\today}

\pacs{12.60.Cn, 12.15Lk}

\author{Tomohiro Abe}
\affiliation{
  Department of Physics, Nagoya University, Nagoya 464-8602, Japan
}
\author{Shinya Matsuzaki}
\affiliation{
  Department of Physics and Astronomy, 
  University of North Carolina, Chapel Hill, NC 27599, USA
}
\author{Masaharu Tanabashi}
\affiliation{
  Department of Physics, Nagoya University, Nagoya 464-8602, Japan
}

\begin{abstract}
We complete the list of one loop renormalization
group equations and matching conditions relevant for the
computation of the electroweak precision parameters $S$ and $T$
in the three site Higgsless model.
We obtain one-loop formulas for 
$S$ and $T$ expressed in terms of physical observables such as the KK
gauge boson mass $M_{W'}$, the KK fermion mass $M$, and the KK gauge
boson ($W'$) 
couplings with light quarks and leptons $g_{W'ff}$.
It is shown that these physical observables, $M_{W'}$, $M$ and
$g_{W'ff}$ are 
severely constrained by the electroweak precision data.
Unlike the tree level analysis on the ideally delocalized fermion, 
we find that perfect fermiophobity of $W'$ is ruled out by
the precision data.
We also study the cutoff dependence of our analysis.
Although the model is non-renormalizable, the dependence on
the cutoff parameter $\Lambda$ is shown to be non-significant.

\end{abstract}

\maketitle

\section{Introduction}
Does a scalar Higgs boson necessarily exist as required in
the standard model?
Higgsless models\cite{Csaki:2003dt} provide a negative answer to this
question, 
achieving the electroweak symmetry breaking (EWSB) without invoking an 
elementary Higgs particle\cite{Higgs:1964ia}.  
The unitarity of longitudinally polarized $W$ and $Z$ boson scattering
is effectively maintained beyond the TeV energy scale by the exchange
of an infinite tower of massive spin-1 
particles\cite{SekharChivukula:2001hz,Chivukula:2002ej,Chivukula:2003kq,He:2004zr},
rather than a spinless Higgs boson. 
These massive spin-1 particles may be provided as Kaluza-Klein (KK)
modes ($W'$, $W''$, $W'''$ $\cdots$, $Z'$, $Z''$, $Z'''$ $\cdots$) of
five dimensional gauge fields compactified on a TeV scale 
interval with appropriate boundary
conditions\cite{Agashe:2003zs,Csaki:2003zu,Burdman:2003ya,Cacciapaglia:2004jz}.  
Deconstruction\cite{ArkaniHamed:2001ca,Hill:2000mu} of extra
dimensions enables us to construct four 
dimensional gauge invariant phenomenological models which approximate
the five dimensional Higgsless theories.
The deconstructed Higgsless
models\cite{Foadi:2003xa,Hirn:2004ze,Casalbuoni:2004id,Chivukula:2004pk,Perelstein:2004sc,Georgi:2004iy,SekharChivukula:2004mu}
(or linear Moose\cite{Georgi:1985hf} models) can be 
used as tools to compute the low energy properties of Higgsless
theories below the cutoff scale. 


Higgsless models are often regarded as ``dual''\cite{Maldacena:1997re,Gubser:1998bc,Witten:1998qj,Aharony:1999ti} to models of dynamical 
EWSB\cite{Weinberg:1979bn,Susskind:1978ms}
such as ``walking technicolor''\cite{Holdom:1981rm,Holdom:1984sk,Yamawaki:1985zg,Appelquist:1986an,Appelquist:1986tr,Appelquist:1987fc}.
The deconstructed version of the Higgsless model possesses extended
electroweak gauge symmetries,
which can be thought as analogues of hidden local symmetries
(HLS)\cite{Bando:1984ej,Bando:1985rf,Bando:1987ym,Bando:1987br,Tanabashi:1993sr,Harada:2003jx} 
and the vector limit symmetry\cite{Georgi:1989xy}
in the dynamical chiral symmetry breaking of QCD\@. 
The gauge sector of the Higgsless model therefore has some
similarities to the BESS
models\cite{Casalbuoni:1985kq,Casalbuoni:1995qt}.

Any phenomenologically viable EWSB model should satisfy constraints
from the precision electroweak
parameters\cite{Peskin:1991sw,Altarelli:1990zd,Altarelli:1991fk}.
An analysis in a very general class of linear Moose model shows that a
Higgsless model with brane localized fermion cannot 
satisfy simultaneously both unitarity bounds and precision electroweak 
constraints at tree level.\cite{SekharChivukula:2004mu} 
Delocalizing fermions within the extra dimension may reduce
significantly the sizes
of electroweak
corrections.\cite{Cacciapaglia:2004rb,Cacciapaglia:2005pa,Foadi:2004ps,Foadi:2005hz}  
In the deconstruction language the delocalizing fermions may be
achieved by allowing fermions to derive electroweak properties from
more than one site over the deconstruction
lattice.\cite{Chivukula:2005bn,Casalbuoni:2005rs,SekharChivukula:2005xm} 
It has been shown that, for an arbitrary Higgsless model with
``ideally'' delocalized fermions\cite{SekharChivukula:2005xm}, three
$(\hat{S}, \hat{T}, W)$ of the 
leading zero-momentum precision electroweak parameters defined by
Barbieri {\it et al.}\cite{Barbieri:2004qk,Chivukula:2004af} vanish at
tree level. 
Moreover, the ideal delocalization implies fermiophobic $W'$.
Presently existing direct $W'$ searches cannot be applied for such a
fermiophobic $W'$ boson. 
The strongest constraints on the $W'$ and $Z'$ masses in the Higgsless 
models then come from the limits on deviations in multi-gauge-boson
vertices.\cite{Chivukula:2005ji,SekharChivukula:2005cc}

Recently, a three site Higgsless model\cite{SekharChivukula:2006cg}
has been proposed. 
This theory incorporates only three sites on the deconstruction
lattice, 
and may be regarded as the simplest deconstructed Higgsless model.
The three site model possesses the same gauge group structure as that 
of the BESS model\cite{Casalbuoni:1985kq,Casalbuoni:1995qt}.
The new interest is aroused in its fermion sector such as   the
fermion delocalization.
The fermionic loop correction to the $\rho$ parameter ($T$ parameter)
has been calculated in the Higgsless 
model.\cite{SekharChivukula:2006cg,Coleppa:2006fu} 
With the aid of electroweak chiral
Lagrangian\cite{Appelquist:1980ae,Appelquist:1980vg,Longhitano:1980iz,Longhitano:1980tm,Appelquist:1993ka,Herrero:1993nc}, 
and using the technique described in \cite{Tanabashi:1993sr,Harada:2003jx}, 
the bosonic chiral logarithmic loop
corrections\cite{Gasser:1983yg,Gasser:1984gg}
to the electroweak precision
parameters have been evaluated 
in \cite{Matsuzaki:2006wn,Sekhar Chivukula:2007ic,Dawson:2007yk,Barbieri:2008cc}.   

In this paper, 
we further investigate the structure of one loop radiative corrections
in the three site Higgsless model, and present the complete list of one loop
renormalization group equations and matching conditions relevant for
the computation of $S$ and $T$ including all of fermionic- and 
bosonic-loop corrections. 
We obtain one-loop formulas $S$ and $T$ expressed  in terms of
physical observables such as the KK 
gauge boson mass $M_{W'}$, the KK fermion mass $M$, and the $W'$
couplings with light quarks and leptons $g_{W'ff}$.
It is shown that the $W'ff$ couplings and the KK fermion mass are
severely constrained by the electroweak precision parameters. 
Unlike the tree level analysis on the ideally delocalized fermion, we
find that the $W'$ boson needs to have non-vanishing coupling with
light fermions in order to satisfy the precision electroweak
constraints.  
The dependence on the ultraviolet cutoff parameter is also studied.
In spite of the non-renormalizability of the model,
we find the cutoff dependence is not significant in the numerical
analysis. 


The next section of the paper introduces the gauge sector Lagrangian,
the symmetry breaking sector Lagrangian, and the fermion sector Lagrangian.
We present the fermion sector Lagrangian above the KK fermion mass
scale $M$ and obtain the effective fermion sector Lagrangian after
integrating out the KK fermion fields below $M$. 
A new operator $x_1'$, which affects the on-shell $W'ff$ coupling strength,
needs to be introduced to absorb the one loop divergence.
In Section III, we summarize the renormalization group equations
relevant for the computation of $S$ and $T$ parameters.
Section IV is for the matching conditions to the electroweak chiral
Lagrangian.
Section V then gives our final results on the $S$ and $T$ parameters
expressed in terms of physical observables.
We then compare our formulas with the precision electroweak data  in
Section VI, 
and find that the $W'ff$ couplings and the KK fermion mass are severely
bounded.
Especially, it is shown that the precision electroweak data reject the
perfect fermiophobity of the $W'$ boson.
Section VII summarizes the results of our analysis.
The renormalization group equation of the operator $x_1'$ is derived
in the Appendix A.

\section{Three site Higgsless model}

In this section we make a brief review on the three site Higgsless
model\cite{SekharChivukula:2006cg}.  
Higher order counter terms, as well as the lowest order Lagrangian
terms, are introduced so as to subtract one-loop divergences.

\begin{figure}[htbp]
  \centering
  \includegraphics[width=6cm]{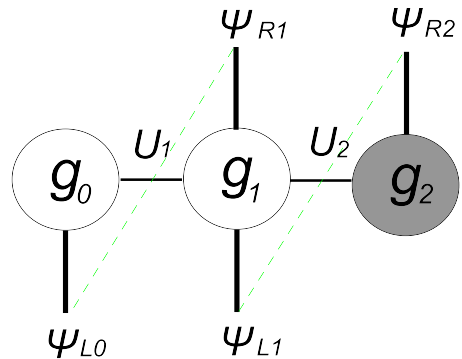}
  \caption{
The three site Higgsless Moose diagram.
Sites 0 and 1 represent $SU(2)$ gauge fields with coupling
strengths $g_0$ and $g_1$.  
Site 2 corresponds to a $U(1)$ gauge field with coupling strength $g_2$.
The non-linear sigma model field $U_1$ ($U_2$) is denoted by link
between sites 0 and 1 (sites 1 and 2).
The left-handed fermions $\psi_{L0}$ and $\psi_{L1}$ feel the gauge
fields at sites 0 and 1, while 
the right-handed fermions $\psi_{R1}$ and $\psi_{R2}$ at sites 1 and
2.
The dashed green lines represent corresponding Yukawa couplings.
}
  \label{fig:moose}
\end{figure}

\subsection{Gauge symmetry and its breaking}

We first consider the gauge sector Lagrangian in the three site
Higgsless model. 
The Moose diagram of this model is illustrated in Figure~\ref{fig:moose}. 
The model incorporates $SU(2)\times SU(2) \times U(1)$ gauge fields
$V_{i\mu}$ ($i=0,1,2$) and two $SU(2)$ non-linear sigma model fields
$U_i$ ($i=1,2$). 
The VEVs $\VEV{U_1}$ and $\VEV{U_2}$ break the gauge symmetry down to
$U(1)_{em}$. 
The model therefore possesses $W'$ and $Z'$ bosons, identified as the
KK gauge bosons,  in addition to the usual $W$, $Z$ and photon.
The corresponding lowest order Lagrangian ${\cal L}_2$ is given by 
\begin{equation}
  {\cal L}_2 = 
    \sum_{i=1}^{2} \dfrac{f_i^2}{4} 
    \tr\left[(D_\mu U_i)^\dagger (D^\mu U_i) \right]
   -\sum_{i=0}^{2} \dfrac{1}{2g_i^2}
    \tr\left[V_{i\mu\nu}V_i^{\mu\nu} \right],
\label{eq:p2}
\end{equation}
where $D_\mu U_i$ and $V_{i\mu\nu}$ are defined as
\begin{equation}
  D_\mu U_i \equiv \partial_\mu U_i + i V_{(i-1)\mu} U_i
             - i U_i V_{i\mu},
\end{equation}
\begin{equation}
  V_{i\mu\nu} \equiv \partial_\mu V_{i\nu} - \partial_\nu V_{i\mu}
               +i[V_{i\mu}, V_{i\nu}].
\end{equation}
The first two gauge fields ($i=0,1$) correspond to $SU(2)$ gauge groups,
\begin{equation}
  V_{i\mu} = \sum_{a=1,2,3} \frac{\tau^a}{2} V^a_{i\mu}, \qquad
  \mbox{for $i=0,1$},
\end{equation}
while the last gauge field ($i=2$) corresponds to $U(1)$ group
embedded as the $T_3$ generator of $SU(2)$,
\begin{equation}
  V_{2\mu} = \frac{\tau^3}{2} V^3_{2\mu}, \qquad
  \mbox{for $i=2$}. 
\label{eq:viosu2}
\end{equation}
The violation of the global $SU(2)$ invariance in Eq.(\ref{eq:viosu2}) 
causes an $SU(2)$ violating one-loop divergence.
We subtract this divergence by introducing a counter 
term\cite{Sekhar Chivukula:2007ic},
\begin{equation}
  {\cal L}_2' = \beta_{(2)} \dfrac{f_2^2}{4} 
    \tr\left[ U_2^\dagger (D_\mu U_2)\tau^3 \right]
    \tr\left[ U_2^\dagger (D^\mu U_2)\tau^3 \right].
\label{eq:beta2}
\end{equation}
The counter term Eq.(\ref{eq:beta2}) is of ${\cal O}(p^2)$ in
the usual chiral perturbation order
counting rule\cite{Gasser:1983yg,Gasser:1984gg}.   
This situation is similar to the custodial $SU(2)$ symmetry violation
in the electroweak chiral Lagrangian.\cite{Appelquist:1980ae,Appelquist:1980vg,Longhitano:1980iz,Longhitano:1980tm,Appelquist:1993ka,Herrero:1993nc}.  

Other one-loop divergences arising from the lowest order Lagrangian
Eq.(\ref{eq:p2}) are invariant under the global $SU(2)$ symmetry.
In order to subtract these $SU(2)$ invariant one-loop divergences, we 
introduce appropriate counter 
terms\cite{Sekhar Chivukula:2007ic}\footnote{
We assume here an approximate global $SU(2)$ symmetry at the cutoff
scale.
See Ref.\cite{Sekhar Chivukula:2007ic} 
for a complete list of ${\cal O}(p^4)$ counter terms including  
$SU(2)$ violating operators.
}
\begin{equation}
  {\cal L}_4 = \sum_{i=1}^2 {\cal L}_{4}^{(i)},
\end{equation}
with
\begin{eqnarray}
  {\cal L}_{4}^{(i)} 
  &=& \alpha_{(i)1} \tr\left[ 
        V_{(i-1)\mu\nu} U_i V_{i}^{\mu\nu} U_i^\dagger
      \right]
  \nonumber\\
  & & -2i\alpha_{(i)2} \tr\left[
        (D_\mu U_i)^\dagger (D_\nu U_i) V_{i}^{\mu\nu}
      \right]
  \nonumber\\
  & & -2i\alpha_{(i)3} \tr\left[
        V_{(i-1)}^{\mu\nu} (D_\mu U_i) (D_\nu U_i)^\dagger
      \right] 
  \nonumber\\
  & & +\alpha_{(i)4} 
       \tr\left[(D_\mu U_i) (D_\nu U_i)^\dagger \right] 
       \tr\left[(D^\mu U_i) (D^\nu U_i)^\dagger \right] 
  \nonumber\\
  & & +\alpha_{(i)5} 
       \tr\left[(D_\mu U_i) (D^\mu U_i)^\dagger \right] 
       \tr\left[(D_\nu U_i) (D^\nu U_i)^\dagger \right] .
  \nonumber\\
  & &
\label{eq:p4}
\end{eqnarray}
These terms are of ${\cal O}(p^4)$ in the conventional chiral
perturbation order counting rule.
Again, the situation is similar to the electroweak chiral Lagrangian.  

Reflecting its coarse deconstruction, 
the delay of unitarity violation is shown to be modest in the lowest
order three site model Eq.(\ref{eq:p2}):
the unitarity in the elastic $W_L W_L$ scattering amplitude is
violated at $2$TeV for $f_1=f_2$ with this 
setup.\cite{Sekhar Chivukula:2006we}
It is expected, however, if the higher order operators
Eq.(\ref{eq:p4}) are included in the analysis, unitarity may hold up
to a higher energy scale.

\subsection{Fermion sector above $M$}
We next turn to the fermion sector.
Quantum numbers of quark/lepton fields are summarized in
Table~\ref{tab:quantumnumbers}. 
The superscript $n=1,2,3$ specifies the generation of quarks and
leptons.
The chiralities are denoted by subscript $L$ and $R$.
Fields $q_L^{n,0}$ and $\ell_L^{n,0}$,
denoted as $\psi_{L0}$ in Figure~\ref{fig:moose},
live on the site 0 in the Moose diagram, while $q_{L,R}^{n,1}$ and 
$\ell_{L,R}^{n,1}$,
denoted as $\psi_{L1}$ and $\psi_{R1}$ in Figure~\ref{fig:moose},
are on the site 1.
The site 2 fermion field $\psi_{R2}$ denotes $u_R^{n,2}$, $d_R^{n,2}$ and
$e_R^{n,2}$.  
This model incorporates heavier KK-fermions 
as well as the ordinary light quarks and leptons.
The kinetic term Lagrangian for these fermion fields is given by
\begin{eqnarray}
{\cal L}_{\rm kinetic}
  &=& \sum_{i=0,1} \sum_{n} \bar{q}_L^{n,i} i\fsl{D} q_L^{n,i}
   +\sum_{n} \bar{q}_R^{n,1} i\fsl{D} q_R^{n,1}
  \nonumber\\
  & &
   +\sum_{n} \bar{u}_R^{n,2} i\fsl{D} u_R^{n,2}
   +\sum_{n} \bar{d}_R^{n,2} i\fsl{D} d_R^{n,2}
  \nonumber\\
  & & + \sum_{i=0,1} \sum_{n} \bar{\ell}_L^{n,i} i\fsl{D} \ell_L^{n,i}
   +\sum_{n} \bar{\ell}_R^{n,1} i\fsl{D} \ell_R^{n,1}
  \nonumber\\
  & &
   +\sum_{n} \bar{e}_R^{n,2} i\fsl{D} e_R^{n,2}.
\end{eqnarray}

\begin{table}[htbp]
  \centering
  \begin{tabular}{c|cccc}
            & $SU(3)_c$ & $SU(2)_0$ & $SU(2)_1$ & $U(1)_2$ \\
    \hline
    $q_L^{n,0}$  & \UL{3}   & \UL{2}    & \UL{1}    & $1/6$ \\
    $\ell_L^{n,0}$ & \UL{1}   & \UL{2}    & \UL{1}    & $-1/2$ \\
    $q_L^{n,1}$  & \UL{3}   & \UL{1}    & \UL{2}    & $1/6$ \\
    $\ell_L^{n,1}$ & \UL{1}   & \UL{1}    & \UL{2}    & $-1/2$ \\
    $q_R^{n,1}$  & \UL{3}   & \UL{1}    & \UL{2}    & $1/6$ \\
    $\ell_R^{n,1}$ & \UL{1}   & \UL{1}    & \UL{2}    & $-1/2$ \\
    $u_R^{n,2}$  & \UL{3}   & \UL{1}    & \UL{1}    & $2/3$ \\
    $d_R^{n,2}$  & \UL{3}   & \UL{1}    & \UL{1}    & $-1/3$ \\
    $e_R^{n,2}$  & \UL{1}   & \UL{1}    & \UL{1}    & $-1$
  \end{tabular}
  \caption{
    Quantum numbers of quark/lepton fields.
    The superscript $n$ specifies the generation of quarks and
    leptons.
    The chiralities are denoted by subscripts $L$ and $R$.
 }
  \label{tab:quantumnumbers} 
\end{table}

Mass terms of these fermion fields are assumed to be
\begin{eqnarray}
  {\cal L}_{\rm mass}
  &=& - m_1 \sum_n \bar{q}_L^{n,0} U_1 q_R^{n,1}
    - m_1 \sum_n \bar{\ell}_L^{n,0} U_1 \ell_R^{n,1} 
  \nonumber\\
  & & 
    - M \sum_n \bar{q}_L^{n,1} q_R^{n,1}
    - M \sum_n \bar{\ell}_L^{n,1} \ell_R^{n,1}
    + \mbox{h.c.},
  \nonumber\\
  & & 
\label{eq:massterm1}
\end{eqnarray}
and
\begin{equation}
  {\cal L}_{\rm mass}'
  = 
    - m_t' \bar{q}_L^{3,1} U_2 P_t 
      \left(
      \begin{array}{c}
        t_{R}^2 \\
        b_{R}^2
      \end{array}
      \right)
    +\mbox{h.c.},
    \qquad
    P_t \equiv \left(
      \begin{array}{cc}
        1 & 0 \\
        0 & 0
      \end{array}
      \right),
\label{eq:massterm2}
\end{equation}
with $t_R^2$ and $b_R^2$ 
being defined by
\begin{equation}
  t_R^2 \equiv u_R^{3,2}, \qquad
  b_R^2 \equiv d_R^{3,2}.
\end{equation}
Here, for simplicity, we incorporate Yukawa coupling of $U_2$ only
for the top quark. 
The other light quarks and leptons thus remain massless in this setup.
It is straightforward to obtain realistic masses of quarks and leptons
(and CKM mixings) by allowing more general Yukawa couplings with $U_2$,
however.
We also assumed $m_1$ and $M$ are  $n$-independent (generation
independent).
This ensures the suppression of FCNC through the GIM mechanism even
when we introduce more general $U_2$ Yukawa couplings.
We assumed $m_1$ and $M$ are quark-lepton universal. 
As we will discuss later, the ratio $m_1/M$ affects the $W$ couplings with
ordinary light quarks and leptons.
The universal $m_1/M$ thus guarantees the quark-lepton universality of
the $W$ couplings.

In order to renormalize the one loop divergences arising from the
Yukawa couplings in Eq.(\ref{eq:massterm1}),
we need to introduce
\begin{eqnarray}
  {\cal L}_{f} 
  &=& -x_1 \sum_n \bar{q}_L^{n,0} i\gamma^\mu
           (D_\mu U_1)U_1^\dagger q_L^{n,0}
  \nonumber\\
  & &
      -x_1 \sum_n \bar{\ell}_L^{n,0} i\gamma^\mu
           (D_\mu U_1)U_1^\dagger \ell_L^{n,0}
  \nonumber\\
  & & +\tilde{x}_1 \sum_n \bar{q}_R^{n,1} i\gamma^\mu
           U_1^\dagger (D_\mu U_1) q_R^{n,1}
  \nonumber\\
  & &
      +\tilde{x}_1 \sum_n \bar{\ell}_R^{n,1} i\gamma^\mu
           U_1^\dagger (D_\mu U_1)\ell_R^{n,1}.
\label{eq:f_counter1}
\end{eqnarray}
We also need to introduce
\begin{eqnarray}
  {\cal L}_{f}'
  &=& -x_{2} \bar{q}_L^{3,1} i\gamma^\mu
       (D_\mu U_2)U_2^\dagger q_L^{3,1}
  \nonumber\\
  & &
      -x_{23} \bar{q}_L^{3,1} i\gamma^\mu q_L^{3,1}
       \tr\left[ \tau^3 U_2^\dagger D_\mu U_2\right]
  \nonumber\\
  & & +\tilde{x}_2 \left( \bar{t}_R^2, \bar{b}_R^2 \right) i\gamma^\mu 
  P_t U_2^\dagger (D_\mu U_2) P_t \left(
    \begin{array}{c}
      t_R^2 \\
      b_R^2
    \end{array}
  \right) 
  \nonumber\\
  & & -\Delta M_t \bar{q}_L^{3,1} U_2 P_t U_2^\dagger q_R^{3,1}
  \nonumber\\
  & &
      -\Delta M_t \bar{q}_R^{3,1} U_2 P_t U_2^\dagger q_L^{3,1},
\label{eq:f_counter2}
\end{eqnarray}
for the renormalization of the one loop divergences arising from the
Yukawa coupling in Eq.(\ref{eq:massterm2}).

\subsection{Fermion sector below $M$}

We assume
\begin{equation}
  M \gg m_1, m_t'. 
\end{equation}
The KK fermion mass is well approximated by the Dirac mass $M$ in this
limit.
We see that the KK fermions are largely made from $q_{L,R}^{n,1}$
and $\ell_{L,R}^{n,1}$.
It is easy to integrate out these KK fermion fields
$q_{L,R}^{n,1}$ and $\ell_{L,R}^{n,1}$. 
The low energy effective theory below $M$ is described by
\begin{equation}
  {\cal L}_{\rm mass}
  = - m_t \bar{q}_L^{3,0} U_1 U_2 P_t \left(
    \begin{array}{c}
      t_R^2 \\
      b_R^2
    \end{array}
    \right),
\label{eq:fermion1}
\end{equation}
\begin{eqnarray}
  {\cal L}_{f}
  &=& -\sum_n x_1 \bar{q}_L^{n,0} i (D_\mu U_1) U_1^\dagger \gamma^\mu
                q_L^{n,0}
  \nonumber\\
  & &
    -\sum_n x_1 \bar{\ell}_L^{n,0} i (D_\mu U_1) U_1^\dagger \gamma^\mu
                \ell_L^{n,0},
\label{eq:fermion2}
\end{eqnarray}
and 
\begin{equation}
  {\cal L}_{f}'
  =  \tilde{x}_2 \left( \bar{t}_R^2, \bar{b}_R^2 \right) i\gamma^\mu 
  P_t U_2^\dagger (D_\mu U_2) P_t \left(
    \begin{array}{c}
      t_R^2 \\
      b_R^2
    \end{array}
  \right) 
\label{eq:fermion3}
\end{equation}
in addition to the light fermion kinetic terms and the gauge sector
Lagrangians Eqs.(\ref{eq:p2}), (\ref{eq:beta2}) and (\ref{eq:p4}).

In order to renormalize the one-loop divergences arising from the
fermion Lagrangian Eq.(\ref{eq:fermion2}), we introduce counter terms 
\begin{equation}
  \dfrac{f_{12}^2}{4} \tr\left[
    (D_\mu (U_1 U_2)) (D^\mu (U_1 U_2))^\dagger
  \right],
\end{equation}
and
\begin{eqnarray}
  & &
  - \sum_n x_1' \bar{q}_L^{n,0} 
     [D^\nu (U_1 V_{1\mu\nu} U_1^\dagger)]
     \gamma^\mu q_L^{n,0}
  \nonumber\\
  & &
  - \sum_n x_1' \bar{\ell}_L^{n,0} 
     [D^\nu (U_1 V_{1\mu\nu} U_1^\dagger)]
     \gamma^\mu \ell_L^{n,0}.
\label{eq:x1p}
\end{eqnarray}
At the scale $\mu=M$, these parameters are determined as
\begin{equation}
  f_{12}^2(\mu=M) \simeq 0, 
\end{equation}
and
\begin{equation}
  x_1'(\mu=M) \simeq 0.
\label{eq:x1pbc}
\end{equation}
Note that the operator $x_1'$ leads to non-trivial corrections to the
$W'$ couplings with light fermions at the mass-shell of the $W'$ boson.

\section{Renormalization group equations}

We are now ready to discuss one-loop ultraviolet divergences arising
from the lowest order Lagrangian. 
Similarly to the usual chiral perturbation
theory\cite{Gasser:1983yg,Gasser:1984gg}, 
these divergences can be subtracted through the renormalization of 
higher order counter terms.
The renormalized parameters then depend on a choice
of the renormalization scale $\mu$. 
The renormalization group equations arising from the bosonic chiral
loop corrections have already been reported in 
Ref.\cite{Sekhar Chivukula:2007ic}. 
In this section, we report our newly derived results on 
the fermionic loop diagrams in the three site Higgsless model, as well
as the bosonic loop diagrams.

\subsection{Above KK fermion mass $M$ }

Performing the standard $\overline{\rm MS}$ renormalization, we find
\begin{eqnarray}
  \mu \dfrac{d}{d\mu} f_1^2 
    &=& \dfrac{3}{(4\pi)^2} (g_0^2 + g_1^2) f_1^2
       -\dfrac{1}{2\pi^2} N_g (N_c + 1) m_1^2, 
  \nonumber\\
  & &
  \\
  \mu \dfrac{d}{d\mu} f_2^2 
    &=& \dfrac{3}{(4\pi)^2} (g_1^2 + \frac{1}{2} g_2^2) f_2^2
       -\dfrac{1}{4\pi^2} N_c m_t'^2, 
  \\
  \mu \dfrac{d}{d\mu} \left( \beta_{(2)} f_2^2 \right) 
    &=& \dfrac{3}{4(4\pi)^2} g_2^2 f_2^2,
\label{eq:rge_beta2a}
\end{eqnarray}
for ${\cal O}(p^2)$ parameters, and
\begin{eqnarray}
  \mu \dfrac{d}{d\mu} \left(\dfrac{1}{g_0^2}\right)
  &=& \dfrac{1}{(4\pi)^2} 
      \left[ \dfrac{44}{3} - \dfrac{1}{6} 
                           - \dfrac{2}{3} N_g (N_c + 1)
      \right],
\nonumber\\
  & &
  \\
  \mu \dfrac{d}{d\mu} \left(\dfrac{1}{g_1^2}\right)
  &=& \dfrac{1}{(4\pi)^2} 
      \left[ \dfrac{44}{3} - \dfrac{2}{6}
                           - \dfrac{4}{3} N_g (N_c + 1)
      \right],
\nonumber\\
  & &
  \\
  \mu \dfrac{d}{d\mu} \left(\dfrac{1}{g_2^2}\right)
  &=& \dfrac{1}{(4\pi)^2} 
      \left[ \phantom{\dfrac{44}{3}} - \dfrac{1}{6} 
                           - \dfrac{4}{3} N_g 
                             (\dfrac{13}{18} N_c + \dfrac{5}{2})
      \right],
  \nonumber\\
  & &
  \\
  \mu \dfrac{d}{d\mu} \left( \dfrac{1}{g_s^2} \right)
  &=& \dfrac{1}{(4\pi)^2} \left[ 22 - \dfrac{16}{3} N_g \right] ,
\end{eqnarray}
for gauge coupling strengths.
Here $N_g$ and $N_c$ denote the numbers of generations and colors,
respectively, 
\begin{equation}
  N_g = 3 , \qquad
  N_c = 3, 
\end{equation}
and 
the $SU(3)_c$ gauge coupling is denoted by $g_s$.

We also obtain\cite{Sekhar Chivukula:2007ic} 
\begin{eqnarray}
  \mu \dfrac{d}{d\mu} \alpha_{(i)1} 
  &=& \dfrac{1}{6(4\pi)^2}, 
\label{eq:rge_a1}
  \\
  \mu \dfrac{d}{d\mu} \alpha_{(i)2} 
  &=& \dfrac{1}{12(4\pi)^2}, 
\label{eq:rge_a2}
  \\
  \mu \dfrac{d}{d\mu} \alpha_{(i)3} 
  &=& \dfrac{1}{12(4\pi)^2}, 
\label{eq:rge_a3}
  \\
  \mu \dfrac{d}{d\mu} \alpha_{(i)4} 
  &=& -\dfrac{1}{6(4\pi)^2}, 
\label{eq:rge_a4}
  \\
  \mu \dfrac{d}{d\mu} \alpha_{(i)5} 
  &=& -\dfrac{1}{12(4\pi)^2}, 
\label{eq:rge_a5}
\end{eqnarray}
for ${\cal O}(p^4)$ terms.
We assumed $\beta_{(2)} \ll 1$ in these expression.
Effects of ${\cal O}(\beta_{(2)})$  are suppressed in the RHSs of
these renormalization group equations.
The bosonic loop terms in these expressions have already been reported
in Ref.\cite{Sekhar Chivukula:2007ic}.
The fermionic loop effects, proportional to $N_g$ and/or $N_c$, 
are newly derived in the present study.

The Dirac mass $M$, the Yukawa couplings $m_1$ and $m_t'$ also depend
on the renormalization scale $\mu$.
For the third generation quarks, we find
\begin{eqnarray}
  \mu \dfrac{d}{d\mu} m_1 &=&
  \dfrac{m_1}{(4\pi)^2} \left[
      -8 g_s^2 - \dfrac{1}{6} g_2^2 -3 \dfrac{m_1^2}{f_1^2} 
    \right], 
\label{eq:third_m1}
  \\
  \mu \dfrac{d}{d\mu} M &=&
  \dfrac{M}{(4\pi)^2} \left[
      -8 g_s^2 -\dfrac{9}{2} g_1^2 - \dfrac{1}{6} g_2^2 
      +\dfrac{3}{2} \dfrac{m_1^2}{f_1^2} 
      + \dfrac{m_t'^2}{f_2^2} 
    \right], 
  \nonumber\\
  & &
\label{eq:third_M}
  \\
  \mu \dfrac{d}{d\mu} m_t' &=&
  \dfrac{m_t'}{(4\pi)^2} \left[
      -8 g_s^2 - \dfrac{2}{3} g_2^2
    \right]. 
\label{eq:third_mt}
\end{eqnarray}
Combining Eq.(\ref{eq:third_m1}) and Eq.(\ref{eq:third_M}), we obtain
a renormalization group equation for the ratio $m_1/M$.
\begin{equation}
  \mu\dfrac{d}{d\mu}\left(\dfrac{m_1}{M}\right)
  = \dfrac{1}{(4\pi)^2}\dfrac{m_1}{M} \left(
    \dfrac{9}{2} g_1^2 
        -\dfrac{9}{2}\dfrac{m_1^2}{f_1^2}
        -\dfrac{m_t'^2}{f_2^2}
    \right) .
\label{eq:rge_third_ratio}
\end{equation}

In a similar manner, we find
\begin{eqnarray}
  \mu \dfrac{d}{d\mu} m_1 &=&
  \dfrac{m_1}{(4\pi)^2} \left[
      -8 g_s^2 - \dfrac{1}{6} g_2^2 -3 \dfrac{m_1^2}{f_1^2} 
    \right], 
\label{eq:quark_m1}
  \\
  \mu \dfrac{d}{d\mu} M &=&
  \dfrac{M}{(4\pi)^2} \left[
      -8 g_s^2 -\dfrac{9}{2} g_1^2 - \dfrac{1}{6} g_2^2 
      +\dfrac{3}{2} \dfrac{m_1^2}{f_1^2} 
    \right], 
  \nonumber\\
  & &
\label{eq:quark_M}
\end{eqnarray}
for the first and second generation quark fields, and
\begin{eqnarray}
  \mu \dfrac{d}{d\mu} m_1 &=&
  \dfrac{m_1}{(4\pi)^2} \left[
      - \dfrac{3}{2} g_2^2 -3 \dfrac{m_1^2}{f_1^2} 
    \right], 
\label{eq:lepton_m1}
  \\
  \mu \dfrac{d}{d\mu} M &=&
  \dfrac{M}{(4\pi)^2} \left[
      -\dfrac{9}{2} g_1^2 - \dfrac{3}{2} g_2^2 
      +\dfrac{3}{2} \dfrac{m_1^2}{f_1^2} 
    \right], 
  \nonumber\\
  & &
\label{eq:lepton_M}
\end{eqnarray}
for lepton fields.
Note that the renormalization group equations of the
lepton mass terms Eqs.(\ref{eq:lepton_m1}), (\ref{eq:lepton_M}) 
differ from those of the quark mass terms Eqs.(\ref{eq:quark_m1}),
(\ref{eq:quark_M}).   
The renormalization group equation for the ratio $m_1/M$ 
\begin{equation}
  \mu\dfrac{d}{d\mu}\left(\dfrac{m_1}{M}\right)
  = \dfrac{1}{(4\pi)^2}\dfrac{m_1}{M} \left(
    \dfrac{9}{2} g_1^2 
        -\dfrac{9}{2}\dfrac{m_1^2}{f_1^2}
    \right)
\label{eq:rge_ratio}
\end{equation}
is, on the other hand, universal to all fermions other than the third
generation quarks.  
An implication of this fact will be discussed in
the next subsection.

We next summarize the renormalization group equations for the one-loop
generated operators listed in Eq.(\ref{eq:f_counter1}).
We find
\begin{eqnarray}
  \mu \dfrac{d}{d\mu} x_1 
    &=& -\dfrac{1}{(4\pi)^2} \dfrac{m_1^2}{f_1^2},  
    \\
  \mu \dfrac{d}{d\mu} \tilde{x}_1 
    &=& -\dfrac{1}{(4\pi)^2} \dfrac{m_1^2}{f_1^2}.  
\end{eqnarray}
In a similar manner, the top quark Yukawa coupling $m_t'$ generates 
\begin{eqnarray}
  \mu \dfrac{d}{d\mu} x_2  
    &=& -\dfrac{1}{2(4\pi)^2} 
         \dfrac{m_t'^2}{f_2^2}, 
    \\  
  \mu \dfrac{d}{d\mu} x_{23}  
    &=& -\dfrac{1}{4(4\pi)^2} 
         \dfrac{m_t'^2}{f_2^2}, 
    \\
  \mu \dfrac{d}{d\mu} \tilde{x}_{2}  
    &=& -\dfrac{1}{(4\pi)^2} 
         \dfrac{m_t'^2}{f_2^2}, 
     \\
  \mu \dfrac{d}{d\mu}(\Delta M_t)
    &=& -\dfrac{M}{2(4\pi)^2} \dfrac{m_t'^2}{f_2^2}, 
\end{eqnarray}
at the one loop level.

\subsection{Matching at the KK fermion mass scale $M$}

The effective theory parameters in
Eqs.(\ref{eq:fermion1})--(\ref{eq:fermion3})  
can be determined through the matching conditions at the KK fermion
mass scale $M$,
\begin{equation}
  x_1(\mu=M-0) \simeq 
  x_1(\mu=M+0) + \left. \dfrac{m_1^2}{M^2} \right|_{\mu=M} ,
\label{eq:match_x1}
\end{equation}
\begin{equation}
  m_t(\mu=M) \simeq \left. \dfrac{m_1 m_t'}{M} \right|_{\mu=M}, 
\label{eq:match_mt}
\end{equation}
and
\begin{equation}
  \tilde{x}_2(\mu=M-0)
  \simeq \tilde{x}_2(\mu=M+0)
    + \left. \dfrac{m_t'^2}{M^2} \right|_{\mu=M} , 
\end{equation}
at the tree level.
Higher order corrections to Eqs.(\ref{eq:match_x1}) and
(\ref{eq:match_mt}) are suppressed by $1/(4\pi)^2$ or 
$m_1^2/M^2$ and thus irrelevant in the evaluation of $\alpha S$
and $\alpha T$ at the ${\cal O}(10^{-3})$ level.

Note that the operator $x_1$ affects the $W$ and $Z$ couplings with
light fermions. 
In order to guarantee universal $W$ and $Z$ couplings with light
quarks and leptons, we thus need to arrange $m_1/M$  universal to
fermion flavors in Eq.(\ref{eq:match_x1}).  
As we stressed in the previous subsection, 
the same renormalization group equation Eq.(\ref{eq:rge_ratio}) is
applied for the ratio $m_1/M$ to all fermion flavors other than the
third generation quarks. 
The universality of the weak gauge boson couplings is therefore
ensured once we assume universal $m_1$ and $M$ at the cutoff scale.
The weak gauge boson couplings with the third generation quarks such
as the $Zb\bar{b}$ vertex, on the other hand, may be affected by the
effects coming from the renormalization group equation
Eq.(\ref{eq:rge_third_ratio}).  
We will study this effect in a separated publication.

Loop level corrections to the matching conditions of 
$f_{1,2}^2$ and $\alpha_{(i)j}$ are also suppressed by $1/(4\pi)^2$,
which can be safely neglected in the evaluation of $\alpha S$
and $\alpha T$.
It is easy to see that the matching conditions of $f_{1,2}^2$ and
$\alpha_{(i)j}$ are trivial at the tree level.

The matching condition of $\beta_{(2)}$ needs to be treated more
carefully.
The one-loop corrected matching condition of $\beta_{(2)}$ is 
\begin{equation}
  \left. \beta_{(2)} f_2^2  \right|_{\mu=M-0}
  = \left. \beta_{(2)} f_2^2  \right|_{\mu=M+0}
   +\left. \dfrac{N_c}{24\pi^2} \dfrac{m_t'^4}{M^2} \right|_{\mu=M} .
\label{eq:matching_beta}
\end{equation}
The non-trivial correction arises from the KK top-quark loop
diagram\cite{SekharChivukula:2006cg}. 
As we will see later, the large value of $m_t'$ makes the one loop 
correction 
\begin{equation}
  \dfrac{N_c}{24\pi^2} \dfrac{m_t'^4}{M^2} 
\end{equation}
important in the evaluation of $\alpha T$ at the ${\cal O}(10^{-3})$
level.

\subsection{Below KK fermion mass $M$}

Below the KK fermion mass scale $M$, the third generation quark fields
couple with $U_1 U_2$ through Eq.(\ref{eq:fermion1}), which causes
non-trivial running of $f_{12}^2$ as
\begin{equation}
  \mu \dfrac{d}{d\mu} f_{12}^2 = -\dfrac{N_c}{4\pi^2} m_t^2, 
\label{eq:rge_f12}
\end{equation}
where we have neglected terms proportional to $f_{12}^2$ in the RHS of 
Eq.(\ref{eq:rge_f12}). 
This is justified so long as $f_{12}^2 \ll m_t^2$.
We also find
\begin{equation}
  \mu \dfrac{d}{d\mu} x_1' = -\dfrac{2}{3(4\pi)^2} \dfrac{x_1}{f_1^2}.
\label{eq:rge_x1p}
\end{equation}
See Appendix.\ref{sec:rge_x1p} for a derivation of Eq.(\ref{eq:rge_x1p}).

The renormalization group equations of other ${\cal O}(p^2)$
coefficients are 
\begin{eqnarray}
  \mu \dfrac{d}{d\mu} f_1^2 
    &=& \dfrac{3}{(4\pi)^2} (g_0^2 + g_1^2) f_1^2,
  \\
  \mu \dfrac{d}{d\mu} f_2^2 
    &=& \dfrac{3}{(4\pi)^2} (g_1^2 + \frac{1}{2} g_2^2) f_2^2,
  \\
  \mu \dfrac{d}{d\mu} \left( \beta_{(2)} f_2^2 \right) 
    &=& \dfrac{3}{4(4\pi)^2} g_2^2 f_2^2.
\label{eq:rge_beta2b}
\end{eqnarray}
We also find the renormalization group equations for the gauge
coupling strengths,
\begin{eqnarray}
  \mu \dfrac{d}{d\mu} \left(\dfrac{1}{g_0^2}\right)
  &=& \dfrac{1}{(4\pi)^2} 
      \left[ \dfrac{44}{3} - \dfrac{1}{6} 
                           - \dfrac{2}{3} N_g (N_c + 1)
      \right],
\nonumber\\
  & &
\label{eq:rge_g0}
  \\
  \mu \dfrac{d}{d\mu} \left(\dfrac{1}{g_1^2}\right)
  &=& \dfrac{1}{(4\pi)^2} 
      \left[ \dfrac{44}{3} - \dfrac{2}{6}
      \right],
  \\
  \mu \dfrac{d}{d\mu} \left(\dfrac{1}{g_2^2}\right)
  &=& \dfrac{1}{(4\pi)^2} 
      \left[ \phantom{\dfrac{44}{3}} - \dfrac{1}{6} 
                           - \dfrac{4}{3} N_g 
                             (\dfrac{11}{18} N_c + \dfrac{3}{2})
      \right],
  \nonumber\\
  & &
  \\
  \mu \dfrac{d}{d\mu} \left( \dfrac{1}{g_s^2} \right)
  &=& \dfrac{1}{(4\pi)^2} \left[ 22 - \dfrac{8}{3} N_g \right] .
\end{eqnarray}
The renormalization group equations for $\alpha_{(i)j}$ are 
unchanged, i.e.,  
Eqs.(\ref{eq:rge_a1})---(\ref{eq:rge_a5}) are valid below $M$ 
as long as $f_{12}^2 \ll
f_1^2, f_2^2$.
The renormalization group equation for the delocalization operator
$x_1$ is\cite{Sekhar Chivukula:2007ic}
\begin{equation}
  \mu \dfrac{d}{d\mu} x_1 = \dfrac{3g_1^2}{(4\pi)^2} x_1.
\end{equation}

\section{Electroweak chiral Lagrangian}

Below the $W'$ mass scale, phenomenology of the three site Higgsless
model can be described by the two-site model, i.e., the electroweak chiral
Lagrangian\cite{Appelquist:1980ae,Appelquist:1980vg,Longhitano:1980iz,Longhitano:1980tm,Appelquist:1993ka,Herrero:1993nc},   
\begin{eqnarray}
  {\cal L}_2 &=& 
    \dfrac{f^2}{4} 
    \tr\left[(D_\mu U)^\dagger (D^\mu U) \right]
   -\dfrac{1}{2g_W^2}
    \tr\left[W_{\mu\nu}W^{\mu\nu} \right]
   \nonumber\\
   & &
   -\dfrac{1}{2g_Y^2}
    \tr\left[B_{\mu\nu}B^{\mu\nu} \right].
\end{eqnarray}
where
\begin{equation}
  D_\mu U = \partial_\mu U + i W_{\mu} U
             - i U B_{\mu}, \qquad
  U \equiv U_1 U_2 .
\end{equation}
There also exists a custodial symmetry violating ${\cal O}(p^2)$
operator 
\begin{equation}
  {\cal L}_2' = \beta \dfrac{f^2}{4} 
    \tr\left[ U^\dagger (D_\mu U)\tau^3 \right]
    \tr\left[ U^\dagger (D^\mu U)\tau^3 \right].
\end{equation}
${\cal O}(p^4)$ operators
\begin{eqnarray}
  {\cal L}_{4}
  &=& \alpha_{1} \tr\left[ 
        W_{\mu\nu} U B^{\mu\nu} U^\dagger
      \right]
  \nonumber\\
  & & -2i\alpha_{2} \tr\left[
        (D_\mu U)^\dagger (D_\nu U) B_{\mu\nu}
      \right]
  \nonumber\\
  & & -2i\alpha_{3} \tr\left[
        W^{\mu\nu} (D_\mu U) (D_\nu U)^\dagger
      \right] 
  \nonumber\\
  & & +\alpha_{4} 
       \tr\left[(D_\mu U) (D_\nu U)^\dagger \right] 
       \tr\left[(D^\mu U) (D^\nu U)^\dagger \right] 
  \nonumber\\
  & & +\alpha_{5} 
       \tr\left[(D_\mu U) (D^\mu U)^\dagger \right] 
       \tr\left[(D_\nu U) (D^\nu U)^\dagger \right] 
  \nonumber\\
  & &
\end{eqnarray}
are also introduced in the two-site model.

Using the technique described in Ref.\cite{Sekhar Chivukula:2007ic}, 
we find matching conditions,
\begin{eqnarray}
  f^2 &\simeq& \dfrac{f_1^2 f_2^2}{f_1^2 + f_2^2}+ f_{12}^2 ,
\label{eq:match_f2}
  \\
  \beta f^2 &\simeq& 
  \beta_{(2)} \dfrac{f_1^4 f_2^2}{(f_1^2 + f_2^2)^2}
  ,
\label{eq:betam}
  \\
  \dfrac{1}{g_W^2} 
  &\simeq& \dfrac{1}{g_0^2}, 
  \\
  \dfrac{1}{g_Y^2} 
  &\simeq& \dfrac{1}{g_2^2},
\end{eqnarray}
and
\begin{eqnarray}
  \alpha_{1}
  &\simeq& -\dfrac{1}{g_1^2} 
       \dfrac{f_1^2 f_2^2}{(f_1^2+f_2^2)^2}
      +\left(\alpha_{(1)1}+\dfrac{x_1}{g_0^2}\right)
       \dfrac{f_2^2}{f_1^2+f_2^2}
  \nonumber\\
  & & 
      +\alpha_{(2)1}
       \dfrac{f_1^2}{f_1^2+f_2^2}
  ,
\label{eq:alpha1m}
\end{eqnarray}
at the scale
\begin{equation}
  \mu = M_{W'} \simeq g_1 \dfrac{\sqrt{f_1^2+f_2^2}}{2} .
\end{equation}

We next consider the renormalization group flow in the two-site model.
The renormalization group equations of the two-site model are given by
\begin{eqnarray}
  \mu \dfrac{d}{d\mu} f^2 
    &=& \dfrac{3}{(4\pi)^2} (g_W^2 + \frac{1}{2} g_Y^2) f^2
       -\dfrac{1}{4\pi^2} N_c m_t^2, 
    \nonumber\\ 
    & &
    \\
  \mu \dfrac{d}{d\mu} \left( \beta f^2 \right) 
    &=& \dfrac{3}{4(4\pi)^2} g_Y^2 f^2,
\label{eq:rge2site_beta}
\end{eqnarray}
\begin{eqnarray}
  \mu \dfrac{d}{d\mu} \left(\dfrac{1}{g_W^2}\right)
  &=& \dfrac{1}{(4\pi)^2} 
      \left[ \dfrac{44}{3} - \dfrac{1}{6} -\dfrac{2}{3}N_g (N_c+1) 
      \right],
    \nonumber\\ 
    & &
    \\
  \mu \dfrac{d}{d\mu} \left(\dfrac{1}{g_Y^2}\right)
  &=& \dfrac{1}{(4\pi)^2} 
      \left[ \phantom{\dfrac{44}{3}} - \dfrac{1}{6} 
        -\dfrac{4}{3} N_g (\dfrac{11}{18} N_c + \dfrac{3}{2})
      \right],
    \nonumber\\ 
    & &
\end{eqnarray}
and
\begin{eqnarray}
  \mu \dfrac{d}{d\mu} \alpha_{1} 
  &=& \dfrac{1}{6(4\pi)^2}, 
\label{eq:rge2site_alpha1}
  \\
  \mu \dfrac{d}{d\mu} \alpha_{2} 
  &=& \dfrac{1}{12(4\pi)^2}, 
  \\
  \mu \dfrac{d}{d\mu} \alpha_{3} 
  &=& \dfrac{1}{12(4\pi)^2}, 
  \\
  \mu \dfrac{d}{d\mu} \alpha_{4} 
  &=& -\dfrac{1}{6(4\pi)^2}, 
  \\
  \mu \dfrac{d}{d\mu} \alpha_{5} 
  &=& -\dfrac{1}{12(4\pi)^2}. 
\end{eqnarray}

\section{$S$ and $T$ parameters}

We are now ready to evaluate $S$ and $T$ parameters in the three site
Higgsless model. 
These parameters are defined as the deviations from the standard
model,
\begin{eqnarray}
  \alpha S 
    &\equiv& -16\pi \alpha \left[
               \alpha_1(\mu) - \alpha_1^{\rm SM}(\mu)
             \right], 
\label{eq:alphaS}
    \\
  \alpha T 
    &\equiv& 2 \left[
               \beta(\mu) - \beta^{\rm SM}(\mu)
             \right], 
\label{eq:alphaT}
\end{eqnarray}
with $\alpha_1^{\rm SM}(\mu)$, $\beta^{\rm SM}(\mu)$ being the low
energy chiral coefficients of the ``reference'' standard model with a
``reference'' heavy Higgs boson mass $M_{H,{\rm ref}}$, 
\begin{eqnarray}
  \alpha_1^{\rm SM}(\mu) 
    &=& \dfrac{1}{6(4\pi)^2} \ln\dfrac{\mu}{M_{H,{\rm ref}}},
\label{eq:alpha1SM}
    \\
  \beta^{\rm SM}(\mu)
    &=& \dfrac{3}{4}\dfrac{g_Y^2}{(4\pi)^2} 
        \ln\dfrac{\mu}{M_{H,{\rm ref}}}.
\end{eqnarray}
The running of $\alpha_1$ is
described by Eq.(\ref{eq:rge2site_alpha1}).
We readily find
\begin{equation}
  \alpha_1(\mu) =
  \alpha_1(M_{W'})+\dfrac{1}{6(4\pi)^2}\ln\dfrac{\mu}{M_{W'}}.  
\end{equation}
Combining this expression with Eq.(\ref{eq:alphaS}) and
Eq.(\ref{eq:alpha1SM}), 
we obtain
\begin{equation}
  S = -16\pi \alpha_1(M_{W'})  
        +\dfrac{1}{6\pi}\ln \dfrac{M_{W'}}{M_{H,{\rm ref}}}. 
\end{equation}
We next turn to the $T$ parameter.
Eq.(\ref{eq:rge2site_beta}) reads
\begin{equation}
  \mu \dfrac{d}{d\mu}\beta = \dfrac{3}{4(4\pi)^2}g_Y^2
  - \beta \dfrac{1}{f^2} \mu \dfrac{d}{d\mu} f^2.
\label{eq:beta_rge2}
\end{equation}
Note that
\begin{equation}
  \beta \ll 1, \qquad
  \dfrac{1}{f^2} \mu \dfrac{d}{d\mu} f^2 \ll 1.
\end{equation}
The second term in the RHS of Eq.(\ref{eq:beta_rge2}) can thus be
neglected safely.
We then obtain
\begin{equation}
  \beta(\mu) = \beta(M_{W'}) 
   + \dfrac{3}{4} \dfrac{g_Y^2}{(4\pi)^2} \ln \dfrac{\mu}{M_{W'}},
\end{equation}
and thus
\begin{equation}
  T = \dfrac{2}{\alpha} \beta(\mu=M_{W'})
       -\dfrac{3}{8\pi c^2} \ln \dfrac{M_{W'}}{M_{H,{\rm ref}}}.
\end{equation}


We next use the matching conditions Eq.(\ref{eq:betam}) and
Eq.(\ref{eq:alpha1m}), and obtain
\begin{eqnarray}
  S &=& \left. \dfrac{16\pi \kappa}{(1+\kappa)^2} \left[
          \dfrac{1}{g_1^2} - \dfrac{x_1}{g_0^2} (1+\kappa)
        \right] \right|_{\mu={M_{W'}}}
    \nonumber\\
    & & \left.
       -\dfrac{16\pi}{1+\kappa}\left(
         \kappa \alpha_{(1)1} + \alpha_{(2)1}
        \right) \right|_{\mu={M_{W'}}}
       +\dfrac{1}{6\pi} \ln \dfrac{M_{W'}}{M_{H,{\rm ref}}},
    \nonumber\\
    & &
\label{eq:SMWp}
    \\
  T &=& \left. \dfrac{1}{\alpha} \dfrac{2}{1+\kappa} \beta_{(2)} 
        \right|_{\mu=M_{W'}}
       -\dfrac{3}{8\pi c^2} \ln \dfrac{M_{W'}}{M_{H,{\rm ref}}},
\label{eq:TMWp}
\end{eqnarray}
with $\kappa$ being defined as
\begin{equation}
  \kappa \equiv \left. \dfrac{f_2^2}{f_1^2} \right|_{\mu=M_{W'}}.
\end{equation}
Assuming $f_{12}^2 \ll f_1^2, f_2^2$, Eq.(\ref{eq:match_f2}) leads
\begin{equation}
  \dfrac{1}{f_1^2} \simeq \dfrac{\kappa}{1+\kappa} \dfrac{1}{f^2}, 
  \qquad
  \dfrac{1}{f_2^2} \simeq \dfrac{1}{1+\kappa} \dfrac{1}{f^2} . 
\label{eq:f1f2}
\end{equation}
In Eqs.(\ref{eq:SMWp}) and (\ref{eq:TMWp}), $g_0$, $g_1$, $x_1$,
$\alpha_{(1)1}$ and $\alpha_{(2)1}$ are renormalized at the scale 
\begin{equation}
  \mu = M_{W'} = g_1 \dfrac{\sqrt{f_1^2 + f_2^2}}{2}
      \simeq g_1 f \dfrac{1+\kappa}{2\sqrt{\kappa}},
\end{equation}
with $f$ being the electroweak scale $f\simeq 250$GeV.

It is possible to express these $S$ and $T$ formulas in terms of
the parameters renormalized at the KK fermion mass scale $M$.
Using the renormalization group equations given in Section III.C, we
obtain
\begin{eqnarray}
  S &=& \left. \dfrac{16\pi \kappa}{(1+\kappa)^2} \left[
          \dfrac{1}{g_1^2} - \dfrac{x_1}{g_0^2} (1+\kappa)
        \right] \right|_{\mu={M}}
    \nonumber\\
    & & \left.
       -\dfrac{16\pi}{1+\kappa}\left(
         \kappa \alpha_{(1)1} + \alpha_{(2)1}
        \right) \right|_{\mu={M}}
       +\dfrac{1}{6\pi} \ln \dfrac{M_{W'}}{M_{H,{\rm ref}}}
    \nonumber\\
    & & -\dfrac{1}{6\pi}\ln \dfrac{M_{W'}}{M} 
        + \dfrac{43\kappa} {3(1+\kappa)^2\pi} \ln \dfrac{M_{W'}}{M}
    \nonumber\\
    & &
        -\dfrac{3\kappa}{(1+\kappa)\pi}\dfrac{g_1^2}{g_0^2}
          x_1  \ln \dfrac{M_{W'}}{M},
\label{eq:SMKK}
    \\
  T &=& \left. \dfrac{1}{\alpha} \dfrac{2}{1+\kappa} \beta_{(2)} 
        \right|_{\mu=M-0}
       -\dfrac{3}{8\pi c^2} \ln \dfrac{M_{W'}}{M_{H,{\rm ref}}}
    \nonumber\\
    & & 
       +\dfrac{3}{8\pi(1+\kappa)c^2} \ln \dfrac{M_{W'}}{M} ,
\label{eq:TMKK}
\end{eqnarray}
at the one loop level.
We have neglected running of the weak gauge coupling constant $g_0$ in
these formulas. 
Taking $\kappa=1$, we find
\begin{eqnarray}
  S &=& \left. 4\pi \left[
          \dfrac{1}{g_1^2} - 2\dfrac{x_1}{g_0^2} 
        \right] \right|_{\mu={M}}
    \nonumber\\
    & & \left.
       -8\pi\left(
         \alpha_{(1)1} + \alpha_{(2)1}
        \right) \right|_{\mu={M}}
       +\dfrac{1}{6\pi} \ln \dfrac{M_{W'}}{M_{H,{\rm ref}}}
    \nonumber\\
    & & 
        + \dfrac{41} {12\pi} \ln \dfrac{M_{W'}}{M}
    \nonumber\\
    & &
        -\dfrac{3}{2\pi}\dfrac{g_1^2}{g_0^2}
          x_1  \ln \dfrac{M_{W'}}{M},
\label{eq:SMKK1}
    \\
  T &=& \left. \dfrac{1}{\alpha} \beta_{(2)} 
        \right|_{\mu=M-0}
       -\dfrac{3}{8\pi c^2} \ln \dfrac{M_{W'}}{M_{H,{\rm ref}}}
    \nonumber\\
    & & 
       +\dfrac{3}{16\pi c^2} \ln \dfrac{M_{W'}}{M} .
\label{eq:TMKK1}
\end{eqnarray}

We now compare Eqs.(\ref{eq:SMKK1}) and (\ref{eq:TMKK1}) with
those given in 
Refs.\cite{Matsuzaki:2006wn,Sekhar Chivukula:2007ic}.
Note that the effective theory used in 
Refs.\cite{Matsuzaki:2006wn,Sekhar Chivukula:2007ic}
does not include the KK fermion.
This theory is therefore valid only below the KK fermion mass scale
$M$.
The cutoff scale used in 
Refs.\cite{Matsuzaki:2006wn,Sekhar Chivukula:2007ic}
should then be regarded as the KK fermion mass scale $M$.
It is easy to see that Eqs.(\ref{eq:SMKK1}) and (\ref{eq:TMKK1})
correspond exactly to those found in 
Refs.\cite{Matsuzaki:2006wn,Sekhar Chivukula:2007ic}.  

Note that Eqs.(\ref{eq:SMKK1}) and (\ref{eq:TMKK1}) are written in
terms of the parameters renormalized at the KK fermion mass scale.
These parameters are not directly related with the on-shell
observables, however.
In order to perform a more sensible phenomenological fit, we next express
$S$ and $T$ in terms of on-shell observables.

We first consider the $S$ parameter.
In order to determine the value $x_1(\mu=M_{W'})$ in
Eq.(\ref{eq:SMWp}), we use
\begin{eqnarray}
  g_{W'ff} &=& \left. \dfrac{g_0^2 g_1}{1+\kappa}
  \left[
    -\dfrac{1}{g_1^2} + \dfrac{x_1}{g_0^2}(1+\kappa)
  \right]  \right|_{\mu=M_{W'}}
  \nonumber\\
  & & \qquad
  + g_1 x_1'(M_{W'}) M_{W'}^2, 
\label{eq:Wpff}
\end{eqnarray}
with $g_{W'ff}$ being the on-shell $W'$ couplings with massless quarks
and leptons.  
Here
$x_1'$ is defined in Eq.(\ref{eq:x1p}).
{}From Eq.(\ref{eq:x1pbc}) and Eq.(\ref{eq:rge_x1p}), we see 
that $x_1'$ can be written in terms of $x_1$,
\begin{equation}
  x_1'(\mu=M_{W'}) = -\dfrac{1}{3(4\pi)^2} \dfrac{x_1}{f_1^2} 
    \ln \dfrac{M_{W'}^2}{M^2}.
\label{eq:x1pvalue}
\end{equation}
Note here that the operator $x_1'$ makes
non-trivial correction to the on-shell $g_{W'ff}$ couplings.
Due to the large value of $M_{W'}$, the effect from $x_1'$ turns out
to be numerically significant.

Note that the delocalization coefficient $x_1$ can always be
tuned\cite{Chivukula:2005bn,Anichini:1994xx} 
so as to minimize the $S$ parameter in Eq.(\ref{eq:SMWp}).  
Phenomenologically we know
\begin{equation}
  S \lessim 0.1
\end{equation}
and thus
\begin{equation}
  x_1(M_{W'}) \simeq \left. \dfrac{1}{1+\kappa} \dfrac{g_0^2}{g_1^2}
  \right|_{\mu=M_{W'}} .
\label{eq:x1p_rough}
\end{equation}
This allows us to expand $x_1$ around Eq.(\ref{eq:x1p_rough}).
Eq.(\ref{eq:Wpff}) then gives
\begin{eqnarray}
  \left. \dfrac{x_1}{g_0^2} \right|_{\mu=M_{W'}}
  &=& \left. \dfrac{1}{1+\kappa} \dfrac{1}{g_1^2} 
    \right|_{\mu=M_{W'}}
    + \dfrac{g_{W'ff}}{g_0^2 g_1}
  \nonumber\\
  & &
    + \dfrac{\kappa}{3(1+\kappa)^2(4\pi)^2} \ln \dfrac{M_{W'}^2}{M^2}
    + \cdots .
  \nonumber\\
  & &
\label{eq:x1p_rough2}
\end{eqnarray}
Combining Eqs.(\ref{eq:SMWp}), (\ref{eq:x1pvalue}) and
(\ref{eq:x1p_rough2}), we obtain
\begin{eqnarray}
  S
  &=& -4 \sqrt{\kappa} \dfrac{s^2}{\alpha} 
  \dfrac{M_W}{M_{W'}} \dfrac{g_{W'ff}}{g_W} 
  \nonumber\\
  & &
  - \left. \dfrac{16\pi}{1+\kappa}\left(
    \kappa \alpha_{(1)1} + \alpha_{(2)1}
  \right) \right|_{\mu=M_{W'}}
  \nonumber\\
  & &
  - \dfrac{\kappa^2}{(1+\kappa)^3} \dfrac{1}{3\pi}
    \ln \dfrac{M_{W'}^2}{M^2}
  + \dfrac{1}{12\pi}\ln\dfrac{M_{W'}^2}{M_{H,{\rm ref}}^2},
\nonumber\\
  & &
\label{eq:Spara2}
\end{eqnarray}
with
\begin{equation}
  s^2 \equiv \dfrac{e^2}{g_W^2}.
\end{equation}
Here we neglected $(4\pi)^{-2} g_{W'ff} \ln(M_W/M_{W'})$ corrections.
We also used the tree level formula
\begin{equation}
  \dfrac{g_0}{g_1} = 
  \dfrac{1+\kappa}{\sqrt{\kappa}} \dfrac{M_{W}}{M_{W'}}.
\label{eq:cratio}
\end{equation}
We next use the renormalization group equations Eq.(\ref{eq:rge_a1}) 
in order to evaluate  $\left. \left(
    \kappa \alpha_{(1)1} + \alpha_{(2)1}
  \right) \right|_{\mu=M_{W'}}$
in Eq.(\ref{eq:Spara2}).
We obtain
\begin{eqnarray}
  S
  &=& -4 \sqrt{\kappa} \dfrac{s^2}{\alpha} 
  \dfrac{M_W}{M_{W'}} \dfrac{g_{W'ff}}{g_W} 
  \nonumber\\
  & &
  - \left. \dfrac{16\pi}{1+\kappa}\left(
    \kappa \alpha_{(1)1} + \alpha_{(2)1} 
  \right)\right|_{\mu=\Lambda}
  \nonumber\\
  & &
  - \dfrac{\kappa^2}{(1+\kappa)^3} \dfrac{1}{3\pi}
    \ln \dfrac{M_{W'}^2}{M^2}
  + \dfrac{1}{12\pi}\ln\dfrac{\Lambda^2}{M_{H,{\rm ref}}^2}.
\nonumber\\
  & &
\label{eq:Sfinal}
\end{eqnarray}

We next turn to the $T$ parameter formula Eq.(\ref{eq:TMWp}).
Using the renormalization group equations Eqs.(\ref{eq:rge_beta2a}),
(\ref{eq:rge_beta2b}) and the matching condition
Eq.(\ref{eq:matching_beta}), we find   
$\beta_{(2)}(M_{W'})$ in  Eq.(\ref{eq:TMWp}) can be expressed as
\begin{equation}
  \beta_{(2)}(M_{W'})
  = \beta_{(2)}(\Lambda)
   +\dfrac{3 g_2^2 }{8(4\pi)^2} \ln \dfrac{M_{W'}^2}{\Lambda^2}
   +\left. \dfrac{N_c}{24\pi^2} \dfrac{1}{f_2^2} 
    \dfrac{m_t'^4}{M^2} \right|_{\mu=M} .
\end{equation}
Combining Eq.(\ref{eq:match_x1}) and Eq.(\ref{eq:match_mt}), we see
\begin{equation}
  \left. \dfrac{m_t'^4}{M^2} \right|_{\mu=M}
  = \left. \dfrac{m_t^4}{M^2} \dfrac{1}{x_1^2} \right|_{\mu=M}.
\label{eq:beta2_corr}
\end{equation}
For the one loop evaluation of $\beta_{(2)}$, it is enough to evaluate
$x_1$ and $m_t$ at tree level.
We use the value of $x_1$ renormalized at the
$M_{W'}$ scale 
\begin{equation}
  x_1(M) \simeq x_1(M_{W'}). 
\end{equation}
We thus find
\begin{equation}
  \left. \dfrac{m_t'^4}{M^2} \right|_{\mu=M}
  \simeq 
  (1+\kappa)^2 \dfrac{g_1^4}{g_0^4} \dfrac{m_t^4}{M^2}
  \left[
    1 + (1+\kappa)\dfrac{g_1}{g_0} \dfrac{g_{W'ff}}{g_0}
  \right]^{-2}
, 
\label{eq:beta2_corr2}
\end{equation}
where we used the first two leading terms in Eq.(\ref{eq:x1p_rough2})
for the evaluation of $x_1$. 
Putting Eq.(\ref{eq:cratio}) into Eq.(\ref{eq:beta2_corr2}) we see
\begin{equation}
  \left. \dfrac{m_t'^4}{M^2} \right|_{\mu=M}
  \simeq 
  \dfrac{\kappa^2}{(1+\kappa)^2} \dfrac{M_{W'}^4}{M_W^4} 
  \dfrac{m_t^4}{M^2} 
  \left[
    1 + \sqrt{\kappa}\dfrac{M_{W'}}{M_W} \dfrac{g_{W'ff}}{g_W}
  \right]^{-2}
, 
\end{equation}
and thus
\begin{eqnarray}
  \beta_{(2)}(M_{W'})
  &\simeq& 
   \beta_{(2)}(\Lambda)
  +\dfrac{3\alpha}{32\pi c^2} \ln \dfrac{M_{W'}^2}{\Lambda^2}
  \nonumber\\
  & &
  + \dfrac{N_c}{24\pi^2} \dfrac{\kappa^2}{(1+\kappa)^3} 
    \dfrac{1}{f^2} \dfrac{M_{W'}^4}{M_W^4} 
  \dfrac{m_t^4}{M^2}
  \times
  \nonumber\\
  & & 
  \qquad \times 
  \left[
    1 + \sqrt{\kappa}\dfrac{M_{W'}}{M_W} \dfrac{g_{W'ff}}{g_W}
  \right]^{-2}
 . 
  \nonumber\\
  & & 
\label{eq:beta2MWp}
\end{eqnarray}
Putting Eq.(\ref{eq:beta2MWp}) into Eq.(\ref{eq:TMWp}), we reach an
expression for $T$, 
\begin{eqnarray}
  T &=& \dfrac{2}{\alpha} \dfrac{1}{1+\kappa}\beta_{(2)}(\Lambda)
       - \dfrac{1}{1+\kappa} \dfrac{3}{16\pi c^2} \ln
         \dfrac{\Lambda^2}{M_{W'}^2} 
    \nonumber\\
    & &
       - \dfrac{3}{16\pi c^2} \ln \dfrac{M_{W'}^2}{M_{H,{\rm ref}}^2}
    \nonumber\\
    & & + \dfrac{N_c}{12\pi^2 \alpha} \dfrac{\kappa^2}{(1+\kappa)^4}
          \dfrac{1}{f^2}  \dfrac{M_{W'}^4}{M_W^4} 
  \dfrac{m_t^4}{M^2} \times
   \nonumber\\
   & & \times
  \left[
    1 + \sqrt{\kappa}\dfrac{M_{W'}}{M_W} \dfrac{g_{W'ff}}{g_W}
  \right]^{-2}
 . 
\label{eq:Tfinal}
\end{eqnarray}

\section{Comparison with electroweak fit}

Now we have formulas both for $S$ and $T$ parameters
Eq.(\ref{eq:Sfinal}) and Eq.(\ref{eq:Tfinal}).
In this section, we assume
\begin{equation}
  \kappa=1,
\label{eq:kappa1}
\end{equation}
and
\begin{equation}
  \alpha_{(1)1}(\Lambda)=\alpha_{(2)1}(\Lambda) = 0, \qquad
  \beta_{(2)}(\Lambda)=0,
\label{eq:cutoff_assumption}
\end{equation}
in Eq.(\ref{eq:Sfinal}) and Eq.(\ref{eq:Tfinal}).
The assumption Eq.(\ref{eq:kappa1}) maximizes the scale of unitarity
violation.
Eq.(\ref{eq:cutoff_assumption}) is justified if the physics above the
cutoff scale $\Lambda$ does not affect much to these oblique
parameters $S$ and $T$.
These assumptions lead to 
\begin{eqnarray}
  S
  &=& -4 \dfrac{s^2}{\alpha} 
  \dfrac{M_W}{M_{W'}} \dfrac{g_{W'ff}}{g_W} 
  \nonumber\\
  & &
  - \dfrac{1}{24\pi}
    \ln \dfrac{M_{W'}^2}{M^2}
  + \dfrac{1}{12\pi}\ln\dfrac{\Lambda^2}{M_{H,{\rm ref}}^2} 
  ,
  \\
  T &=& 
       - \dfrac{3}{32\pi c^2} \ln
         \dfrac{\Lambda^2}{M_{W'}^2} 
       - \dfrac{3}{16\pi c^2} \ln \dfrac{M_{W'}^2}{M_{H,{\rm ref}}^2}
    \nonumber\\
    & & + \dfrac{N_c}{192\pi s^2} 
          \dfrac{M_{W'}^4}{M_W^6} 
  \dfrac{m_t^4}{M^2} 
  \left[
    1 + \dfrac{M_{W'}}{M_W} \dfrac{g_{W'ff}}{g_W}
  \right]^{-2}
 . 
 \nonumber\\
 & & 
\end{eqnarray}
Note that these expressions are written in terms of physical
observables, such as $M_{W'}$, $g_{W'ff}$ and the KK fermion mass $M$,
except for the cutoff scale $\Lambda$. 
Note also that the cutoff scale $\Lambda$ should be below the
naive dimensional analysis (NDA)\cite{Manohar:1983md} scale $4\pi f_1
= 4\pi f_2 \simeq 4.3$TeV.

\begin{figure}[htbp]
  \centering
  \includegraphics[width=7cm]{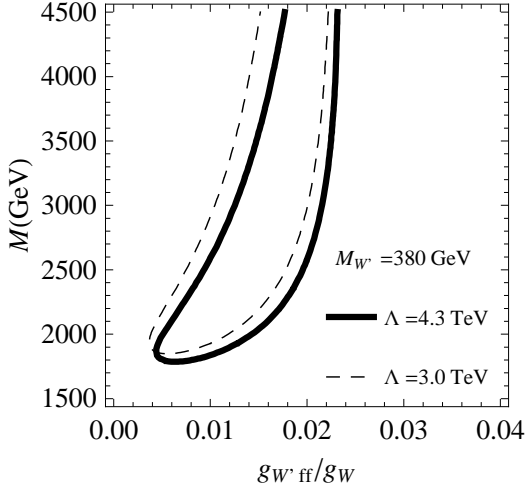}
  \caption{
Phenomenologically acceptable area in the $(g_{W'ff}/g_W, M)$ plane. 
$M_{W'}=380$GeV is assumed.
The region surrounded by the solid (dashed) curve satisfies the 95\% CL
constraint for cutoff $\Lambda=4.3$TeV (3.0TeV).  
}
  \label{fig:43-038}
\end{figure}

\begin{figure}[htbp]
  \centering
  \includegraphics[width=7cm]{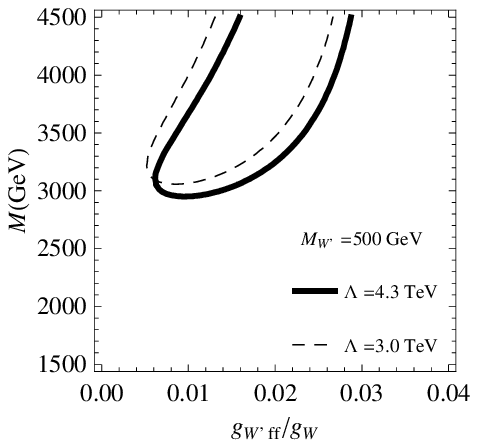}
  \caption{
Phenomenologically acceptable area in the $(g_{W'ff}/g_W, M)$ plane. 
$M_{W'}=500$GeV is assumed.
The region surrounded by the solid (dashed) curve satisfies the 95\% CL
constraint for cutoff $\Lambda=4.3$TeV (3.0TeV).  
}
  \label{fig:43-050}
\end{figure}

\begin{figure}[htbp]
  \centering
  \includegraphics[width=7cm]{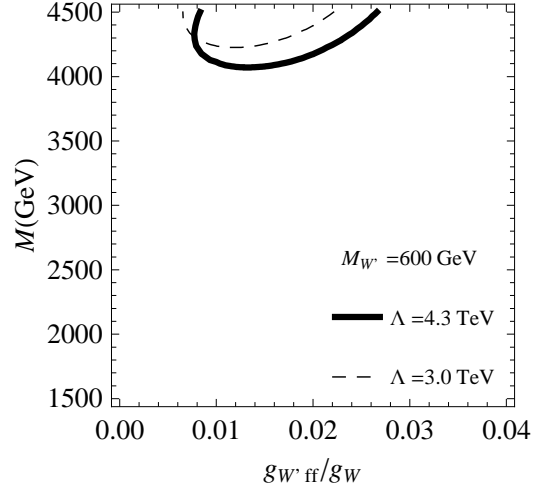}
  \caption{
Phenomenologically acceptable area in the $(g_{W'ff}/g_W, M)$ plane. 
$M_{W'}=600$GeV is assumed.
The region surrounded by the solid (dashed) curve satisfies the 95\% CL
constraint for cutoff $\Lambda=4.3$TeV (3.0TeV).  
}
  \label{fig:43-060}
\end{figure}

We compare these results with values of $S$ and $T$ 
extracted from the precision electroweak fit,
\begin{eqnarray}
  S &=& -0.21 \pm 0.09,
\label{eq:Sfit}
\\
  T &=&  0.21 \pm 0.09.
\label{eq:Tfit}
\end{eqnarray}
The central values of Eqs.(\ref{eq:Sfit}), (\ref{eq:Tfit}) are for
a ``reference'' Higgs boson mass $M_{H,{\rm ref}}=1$TeV, 
and are taken from Figure 10.4 in Ref.\cite{Yao:2006px}.
Note that there exists strong error correlation between $S$ and
$T$,
\begin{equation}
 \rho_{ST}=0.84. 
\end{equation}
The 95\% CL allowed region in the $(S,T)$ plane is then given by
\begin{eqnarray}
& &
  \dfrac{(S-S_0)^2}{\sigma_S^2} 
 +\dfrac{(T-T_0)^2}{\sigma_T^2} 
 -2\rho_{ST} \dfrac{S-S_0}{\sigma_S}\dfrac{T-T_0}{\sigma_T}  
\nonumber \\
& &
 \qquad\qquad < 5.99\times (1-\rho_{ST}^2), 
\label{eq:95percent}
\end{eqnarray}
with $S_0=-0.21$, $T_0=0.21$, $\sigma_S=0.09$, $\sigma_T=0.09$.
For fixed $M_{W'}$ and $\Lambda$, we now obtain bounds
on $g_{W'ff}$ and $M$ from the constraint in the ($S$,$T$) plane. 
The allowed regions satisfying the 95\% CL constraint
Eq.(\ref{eq:95percent}) 
are depicted in Figure~\ref{fig:43-038} in the $(g_{W'ff}/g_W, M)$
plane
for $M_{W'}=380$GeV.
In the plot we used $\alpha^{-1}=128.91$, 
$G_F=1/(\sqrt{2}f^2)=1.16637\times 10^{-5}$GeV$^{-2}$,
$s^2=1-c^2=0.231$, and $m_t=174.2$GeV.

The region surrounded by the solid (dashed) curve satisfies the 95\%
CL constraint Eq.(\ref{eq:95percent}) for cutoff
$\Lambda=4.3$TeV (3.0TeV).
We emphasize that the $W'$ boson needs to have non-zero
$g_{W'ff}$ in order to satisfy the precision electroweak
constraints. 
This is in sharp contrast to the tree level
analysis\cite{SekharChivukula:2006cg}, where the ideal 
delocalization\cite{SekharChivukula:2005xm} implies completely
fermiophobic $W'$. 
The KK fermion mass is also severely constrained.
We find $M\gtrsim 1.8$TeV for $M_{W'}=380$GeV and $\Lambda=4.3$TeV.
The change of the cutoff assumption
to $\Lambda=3.0$TeV affects little in Figure~\ref{fig:43-038}. 
The dependence on the cutoff assumption is 
not significant in this numerical analysis.

\begin{figure}[htbp]
  \centering
  \rotatebox{-90}{\includegraphics[width=5.5cm]{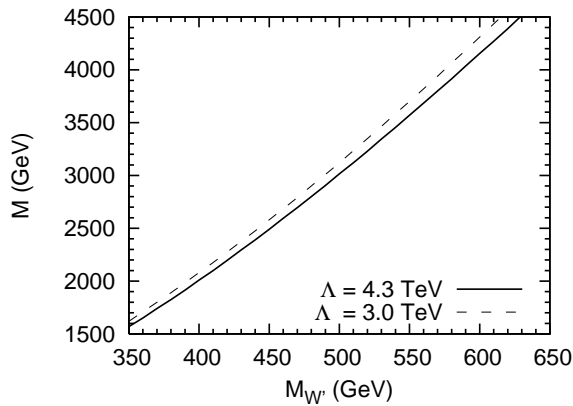}}
  \caption{
Phenomenologically acceptable values of $M_{W'}$ and $M$.
The area above the curve satisfies 95\% CL constraint in the $(S,T)$ plane.
The $W'ff$ coupling $g_{W'ff}$ is assumed to be tuned so as to
maximize the likelihood.
The solid curve is for $\Lambda=4.3$TeV, and dashed for
$\Lambda=3.0$TeV. 
}
  \label{fig:mwp_mf}
\end{figure}

Figures~\ref{fig:43-050} and \ref{fig:43-060} show similar plots for 
$M_{W'}=500$GeV and for $600$GeV.
We see the KK fermion lower mass bound becomes much severer for
heavier 
$M_{W'}$.
This severe constraint on $M$ essentially comes from the $T$
parameter constraint through the KK top-quark loop
effect\cite{SekharChivukula:2006cg}.  
Figure~\ref{fig:mwp_mf} shows the bound on the KK fermion mass $M$ as
functions of $M_{W'}$.  
Here $g_{W'ff}$ is tuned so as to maximize the likelihood in the fit.
Again, we find the cutoff dependence is rather small.


We now turn to the  bounds on $M_{W'}$.
As shown in Ref.\cite{SekharChivukula:2006cg}, the LEP-II triple gauge
vertex bound gives a lower bound on the $W'$ mass
\begin{equation}
  M_{W'} > 380 \mbox{GeV}.
\end{equation}
As depicted in Figure~\ref{fig:mwp_mf},
the lower bound on the KK fermion mass $M$ increases 
monotonically for increasing $M_{W'}$.
In order to keep our effective theory analysis sensible, we
need to require
\begin{equation}
  M < \Lambda .
\label{eq:sensible}
\end{equation}
The requirement Eq.(\ref{eq:sensible}), combined with
Figure~\ref{fig:mwp_mf}, leads to  {\em upper} bounds on the $W'$
boson mass 
\begin{equation}
  M_{W'} < 610 \mbox{GeV}
\end{equation}
for $\Lambda=4.3$TeV, and
\begin{equation}
  M_{W'} < 490 \mbox{GeV}
\end{equation}
for $\Lambda=3.0$TeV.

\section{Conclusions}

Does the three site Higgsless model survive the electroweak precision
tests at loop?
In this paper, we have positively answered to this question,
explicitly showing that there exists a parameter region in which both
$S$ and $T$ are consistent with the present bounds.
We have obtained compact formulas for 
$S$ and $T$ at the one loop level, written in terms of physical
observables such as the KK gauge boson mass $M_{W'}$, the KK fermion
mass $M$, and the $W'$ couplings with light quarks and leptons $g_{W'ff}$.
It has been shown, however, that the $W'ff$ couplings and the KK
fermion mass $M$ are severely constrained.
Especially, unlike the tree level analysis on the ideally delocalized
fermion, it has been shown that perfectly fermiophobic $W'$ is
inconsistent with the electroweak fit.
Small but non-zero value of $g_{W'ff}$ is required.

So far, collider phenomenology of Higgsless $W'$ bosons has been
analyzed assuming perfectly fermiophobic
$W'$~\cite{Birkedal:2004au,He:2007ge}.\footnote{
Very recently, a possibility of non-fermiophobic $W'$ is pointed out  
in the context of four-site Higgsless 
models\cite{Accomando:2008dm,Accomando:2008jh,Chivukula:2008gz}.
}
We found small $g_{W'ff}$ of order $1\times 10^{-2}$ -- $3\times
10^{-2}$ of $g_W$ is required to be consistent with the precision
electroweak data. 
Note that such a $W'$ boson is still effectively fermiophobic at the
Tevatron collider. 
Actually, the production cross section $\sigma(q\bar{q}\to W')$ is
suppressed $1\times 10^{-4}$ --- $1\times 10^{-3}$ compared with a
$W'$ boson having the 
standard model coupling strength.
The existing direct search limit on $W'$ from the
Tevatron\cite{Abulencia:2006kh} cannot be applied for such an
effectively fermiophobic $W'$.
Nevertheless, it should be interesting to investigate the LHC
sensitivity toward such a small $g_{W'ff}$ coupling.

\section*{Acknowledgments}
We thank R. Sekhar Chivukula for his thoughtful suggestions in the 
preparation stage of this manuscript.
T.A. thanks members of particle theory groups at Nagoya University
and Tohoku University for their hospitality during the completion of
this work.
T.A. is supported in part by the JSPS Research Fellowships for Young
Scientists. 
S.M. is supported by the U.S. Department of Energy under Grants
No.~DE-FG02-06ER41418.  
M.T.'s work is supported in part by the JSPS Grant-in-Aid for
Scientific Research No.~20540263.

\appendix

\section{Derivation of Eq.(\ref{eq:rge_x1p})}
\label{sec:rge_x1p}

The gauge field $V_0^\mu$ interacts with the non-linear sigma model
field $U_1$.
Extracting the interaction term from the lowest order chiral
Lagrangian, we find
\begin{equation}
  \left. \dfrac{f_1^2}{4}\tr\left[
    (D_\mu U_1)^\dagger (D^\mu U_1)
  \right] 
  \right|_{V_0^\mu} = - \dfrac{f_1^2}{4} V_0^{a\mu} \tr\left[
    \tau^a i(\partial_\mu U_1)^\dagger U_1
  \right].
\label{eq:app1}
\end{equation}
The delocalization operator Eq.(\ref{eq:fermion2}) 
leads to the Nambu-Goldstone boson interaction with the fermion
currents as 
\begin{eqnarray}
  & & -\sum_n \dfrac{x_1}{2} 
       \bar{q}_L^{n,0} \tau^a \gamma^\mu   q_L^{n,0}
       \tr\left[ \tau^a i (\partial_\mu U_1)^\dagger U_1  \right]
  \nonumber\\
  & &
      -\sum_n \dfrac{x_1}{2} 
       \bar{\ell}_L^{n,0} \tau^a \gamma^\mu  \ell_L^{n,0}
       \tr\left[ \tau^a i (\partial_\mu U_1)^\dagger U_1 \right].
\label{eq:app2}
\end{eqnarray}
Comparing Eq.(\ref{eq:app1}) with Eq.(\ref{eq:app2}), we see 
the delocalization operator $x_1$ can be 
absorbed into the redefinition of $V_{0\mu}^a$,
\begin{equation}
  \tilde{V}_{0\mu}^a \equiv V_{0\mu}^a + 2 \dfrac{x_1}{f_1^2}
  \left( \sum_n \bar{q}_L^{n,0} \gamma_\mu \tau^a q_L^{n,0}
        +\sum_n \bar{\ell}_L^{n,0} \gamma_\mu \tau^a \ell_L^{n,0}
  \right),
\end{equation}
in the Nambu-Goldstone boson field interaction.
In Ref.\cite{Sekhar Chivukula:2007ic}, the authors found that the
Nambu-Goldstone boson loop gives rise to a divergence of
\begin{equation}
  \dfrac{-1}{6(4\pi)^2 \bar{\epsilon}}
  \tr\left[V_{0\mu\nu} U_1 V_1^{\mu\nu} U_1^\dagger \right], \qquad
  \dfrac{1}{\bar{\epsilon}} \equiv
  \dfrac{\Gamma(2-d/2)}{2(4\pi)^{d/2-2}}. 
\label{eq:divergence}
\end{equation}
The divergence factor in Eq.(\ref{eq:divergence}) can also be read
from the renormalization group coefficient in Eq.(\ref{eq:rge_a1}). 
The divergence arising from $x_1$ can be obtained by replacing
\begin{eqnarray}
  V_{0\mu\nu} 
  \to \tilde{V}_{0\mu\nu}
  &=& V_{0\mu\nu} + \dfrac{x_1}{f_1^2} 
     \sum_n \tau^b 
     \left\{
        \partial_\mu(\bar{q}_L^{n,0} \gamma_\nu \tau^b q_L^{n,0})
     \right.
  \nonumber\\
  & & \quad \left.
       -\partial_\nu(\bar{q}_L^{n,0} \gamma_\mu \tau^b q_L^{n,0})
       +\mbox{(lepton part)}
    \right\},
  \nonumber\\
  & &
\end{eqnarray}
in Eq.(\ref{eq:divergence}).
Dropping the total derivative terms, we then find the $x_1$-proportional
part of the divergence,
\begin{eqnarray}
  & & 
  \left. 
  \dfrac{-1}{6(4\pi)^2 \bar{\epsilon}}
  \tr\left[\tilde{V}_{0\mu\nu} U_1 V_1^{\mu\nu} U_1^\dagger 
  \right]
  \right|_{x_1}
  \nonumber\\
  &=& 
    -\dfrac{1}{3(4\pi)^2\bar{\epsilon}} \dfrac{x_1}{f_1^2}
    \sum_n \tr\biggl[
      \partial_\mu (\bar{q}_L^{n,0}\gamma_\nu \tau^b q_L^{n,0}) \tau^b
      U_1 V_1^{\mu\nu} U_1^\dagger  
  \nonumber\\
  & & 
      + \mbox{(lepton part)}
    \biggr]
  \nonumber\\
  &=& -\dfrac{2}{3(4\pi)^2\bar{\epsilon}} \dfrac{x_1}{f_1^2}
    \sum_n \biggl\{
      \bar{q}_L^{n,0}\gamma^\mu D^\nu(U_1 V_{1\mu\nu} U_1^\dagger) 
      q_L^{n,0}
 \nonumber\\
  & &
      + \mbox{(lepton part)}
    \biggr\}.
\end{eqnarray}
It is now straightforward to obtain the renormalization group
equation Eq.(\ref{eq:rge_x1p}).

\end{document}